\documentstyle[sprocl,epsfig]{article}
\def\NPB{{\em Nucl. Phys.} B}
\def\PLB{{\em Phys. Lett.}  B}

\def\ra{\rightarrow}

\def\be{\begin{equation}}
\def\ee{\end{equation}}
\def\bea{\begin{eqnarray}}
\def\eea{\end{eqnarray}}
\newcommand{\half}{\frac{1}{2}}
\newcommand{\tr}{\mbox{Tr}}
\newcommand{\noi}{\noindent}
\newcommand{\en}{\end{equation}}
\newcommand{\eqa}{\begin{eqnarray}}
\newcommand{\ena}{\end{eqnarray}}

\begin{document}
\title{ \vspace*{-2.4cm} {\hspace*{-9cm}\normalsize ITEP-LAT/2002--18} \\[1.5cm]
P-VORTICES, NEXUSES  AND EFFECTS OF GRIBOV  COPIES IN THE CENTER
GAUGES \footnote{Talk given by V.G.B. at the International Symposium on
Quantum Chromodynamics and Color Confinement (Confinement 2000), Osaka,
March 2000.}}
\author{V.G.~Bornyakov}
\address{Institute for High Energy Physics, Protvino 142284, Russia}
\author{D.A.~Komarov, M.I.~Polikarpov, and A.I.~Veselov }
\address{Institute of Theoretical and Experimental  Physics,
B.Cheremushkinskaya 25, Moscow, 117259, Russia}
\maketitle\abstracts{
We perform the careful study of the Gribov  copies problem in $SU(2)$
lattice gauge theory for
maximal direct and maximal indirect center projections.
We find that
this  problem is much more severe than it was thought  before.
The projected string tension
is not in agreement with the physical string tension.
We also show that the particle--like objects, nexuses, might be
important for the confinement dynamics. }

\section{Introduction}
The old idea about the role of the center vortices in confinement 
phenomena~\cite{center_v} has been revived recently with the use of lattice
regularization. Both gauge invariant \cite{tomb1} and gauge dependent
\cite{greens1} approaches are developed. The gauge dependent studies were
done in several center gauges.  The center gauge leaves intact
center group local gauge invariance. It is believed that gauge dependent
P-vortices defined on the lattice plaquettes are able to locate thick gauge
invariant center vortices and thus provide the essential evidence for the
center vortex picture of confinement. So far 3 different center gauges have
been used in practical computations: the indirect maximal center (IMC) gauge
\cite{greens1}, the direct maximal center (DMC) gauge  \cite{greens2} and
the Laplacian center gauge \cite{def}. The first two
suffer from Gribov  copies problem. Many results supporting the important
role of P-vortices are obtained in these two gauges but the
problem of Gribov copies effects, we are addressing here, has not been
studied properly.  We also investigate properties of recently introduced new
objects called nexuses \cite{cornwall,volovik} or center monopoles
\cite{cpvz,bkpv}. One can define nexus in $SU(N)$ gauge theory as a 
particle-like object formed by $N$ center vortices meeting at the center, 
with the zero (mod $N$) net flux. We use P--vortices in the center 
projection to define nexuses in $SU(2)$ lattice gauge theory.

In \cite{bkpv}
we have found that in DMC gauge
the projected string tension is much lower than the physical one
which contradicts the earlier claims. We also confirmed the observation made
in \cite{tomb2}:
there are gauge copies which correspond to higher maxima of
the gauge fixing functional $F$ (see below for definition) than usually
obtained and at the same time these new gauge copies produce P-vortices
evidently with no center vortex finding ability. These results are discussed
in section 2. In section 3 we present our results for IMC gauge. Results of
sections 2 and 3 were obtained with the relaxation - overrelaxation (RO)
algorithm. In section 4 we
present our preliminary results obtained with more effective algorithm,
simulated annealing (SA). We show that the use of this algorithm permits
to obtain higher maxima and thus to solve the puzzle imposed in
\cite{tomb2}. Moreover SA algorithm gives the lowest value of the
string tension. We also discuss the 
finite volume effects~\cite{greens4} for the projected string tension.

\vspace{-0.1cm}
\section{Direct maximal center gauge}

The DMC gauge is defined \cite{greens2} by the maximization of the following
functional:

\vspace{-0.2cm}
\be F(U) =
\frac{1}{4 V} \sum_{n,\mu} \left( \half\tr U_{n,\mu} \right)^2 = \frac{1}{4
       V} \sum_{n,\mu} \frac{1}{4}\left( \tr_{adj} U_{n,\mu} +1 \right),
     \label{maxfunc}
\vspace{-0.0cm}
\en
with respect to local gauge transformations, $U_{n,\mu}$ is the lattice 
gauge field, $V$ is the lattice volume.  DMC gauge condition fixes the gauge 
up to $Z(2)$ gauge transformation.  The fixed configuration can be 
decomposed into $Z(2)$ and coset parts:  $U_{n,\mu} = Z_{n,\mu} V_{n,\mu}$, 
where $Z_{n,\mu} = \mbox{sign} \tr U_{n,\mu}$.  Plaquettes constructed from 
$Z_{n,\mu}$ field have values $\pm 1$. Those of them taking values $-1$ 
compose the so called P-vortices.  P-vortices form closed surfaces in 4D 
space. Some evidence has been collected, that P-vortices in the center  
gauges can serve to locate gauge invariant center vortices. In ref.
\cite{greens2} the projected Wilson loops, $W_{Z(2)}$, are computed via 
linking number of the static quarks trajectories and P-vortices. It was 
found that the string tension, 
$\sigma_{Z(2)}$, obtained from $W_{Z(2)}$ is very close to the physical 
string tension $\sigma_{SU(2)}$. This fact has been called center 
dominance.  Another important observation was that the density of P-vortices 
scales in agreement with asymptotic scaling \cite{greens2,tubing}.  We 
inspect these statements using careful gauge fixing procedure.

The problem of the DMC  gauge fixing is that the functional $F(U)$  has many
local maxima. This is the analogue of the Gribov problem in continuum gauge
theories \cite{gribov}. We call configurations corresponding to different
local maxima Gribov  copies. To perform unambiguous computations one must
fix the gauge completely, i.e. to find the global maxima~\cite{zwanziger}.
It is impossible to do it numerically, and we generate a large  number
$N_{cop}$ of local maxima, calculate observables using configuration 
corresponding to the highest maximum and extrapolate results to $N_{cop} \ra 
\infty$ limit. The local maxima are produced  by applying the RO algorithm 
to random gauge copies of the original configuration.

Our computations have been performed on lattices  $L^4=12^4$ at  $\beta=2.3,
2.4$ and $L^4=16^4$ at $\beta=2.5$.
At  $\beta = 2.3, 2.4$ ($\beta = 2.5$)
we use $100$ ($50$) statistically independent gauge field configurations.
In Table 1 we show  the ratio of string tensions
$\sigma_{Z(2)}/\sigma_{SU(2)}$
\footnote{The data for $\sigma_{SU(2)}$ are taken from \cite{bss}}.
$\sigma_{Z(2)}$ is computed from the Creutz ratio
$\chi (I)$; $3 \leq I \leq 4$
on  $12^4$ lattice, and $3 \leq I \leq 6$ on  $16^4$ lattice.
For $N_{cop}=3$ (the number of gauge copies used in \cite{greens2})
$\sigma_{Z(2)}$ is close to $\sigma_{SU(2)}$. But it becomes significantly
lower for large $N_{cop}$. Thus RO gauge fixing gives the strong
dependence of $\sigma_{Z(2)}$ on $N_{cop}$. In the limit  $N_{cop} \to
\infty$ $\sigma_{Z(2)}$  is  20-30 \% lower than $\sigma_{SU(2)}$.

In Table 1 we also show the ratio $2\rho/\sigma_{SU(2)} a^2$ ($\rho$ is the
density of P-vortices). As it is claimed in ref.  \cite{tubing} $2 \rho$
coincides with the dimensionless string tension, $\sigma_{SU(2)} a^2$ if
plaquettes carrying P-vortices are uncorrelated.  Our results in Table~1
show that the density of P-vortices does not satisfy this relation. We
have found  that for $N_{cop} = 3 \quad \rho$ is in a good agreement with
asymptotic scaling, as it was observed before~\cite{greens2}.  But for
$N_{cop} \to \infty \quad \rho$ deviates from the two loop asymptotic
scaling formula and its dependence on $\beta$ becomes similar to that of
$\sigma_{SU(2)} a^2$.

We also performed computations using the modified gauge fixing procedure
(LRO) suggested in \cite{tomb2}:  every copy  has been first fixed to Landau
gauge, and then the RO algorithm for DMC gauge has been
applied. With this  procedure we found the local maximum higher than local 
maxima of RO procedure for any $N_{cop} \in [1;20]$ and in the limit 
$N_{cop} \to \infty$ (see Table 2). We confirm 
that RO and LRO gauge fixing procedures generate two classes of gauge 
equivalent copies with different properties:  $\sigma_{Z(2)}$ is zero for 
local maxima produced with LRO and nonzero for those produced with RO.  The 
density of P-vortices is essentially lower for LRO.  Then the idea of the 
complete gauge fixing based on the choice of the global maximum would force 
us to choose the LRO local maxima.
Thus the vortex finding property is completely lost. In 
chapter 4 we demonstrate that SA algorithm solves this problem.

\begin{table}[tp]
\caption{The comparison of
$\sigma_{Z(2)}$, $\sigma_{SU(2)}$ and $\rho$ for DMC gauge,  RO gauge fixing
procedure. } \label{t1} \vspace{0.5cm} \setlength{\tabcolsep}{0.55pc}
\begin{centering}
\begin{tabular}{c|ccc|ccc}  \hline $N_{cop}$&
&$\sigma_{Z(2)}/\sigma_{SU(2)}$&&&$2\rho/(\sigma_{SU(2)} a^2)$& \\
\hline &$\beta=2.3$ &$\beta= 2.4$ &$ \beta=
2.5$&$\beta=2.3$&$\beta=2.4$&$\beta=2.5$   \\
\hline 3&0.94(2) &0.93(2) &
0.98(2)     &1.30(1) &1.51(1) & 1.74(1) \\
20&0.87(2)  &0.80(2)  & 0.83(3)
&1.27(1) &1.42(1) & 1.61(2)\\
$\infty$&0.82(3)&0.71(3)& 0.71(3)&1.24(1)
&1.33(2) & 1.49(2) \\
\hline \end{tabular} \end{centering} 
\end{table}

\begin{table}[tp]
\caption{$<F_{max}>$ for DMC gauge obtained with various algorithms}
\label{t1} \vspace{0.5cm}
\setlength{\tabcolsep}{0.55pc}
\begin{centering}
\begin{tabular}{c|ccc}  \hline
&$\beta=2.3$ &$\beta= 2.4$ &$ \beta= 2.5$  \\ \hline
RO($N_{cop}=\infty$)&0.7552(1)&0.7764(2)&0.7943(2)   \\
LRO($N_{cop}=\infty$)&0.7564(1)&0.7759(3)&0.7955(2) \\
SA($N_{cop}=3$) &0.7588(1)&0.7770(2)& 0.7970(2) \\ \hline
\end{tabular} \end{centering} \end{table}

\subsection{Nexuses.}
We also investigate the properties of the point like objects,
called nexuses. On the 4D lattice we have the conserved currents of nexuses,
defined after the center projection. First we calculate the phase, $s_l$, 
of the $Z(2)$ link variable:  $Z_l = \exp(i\pi s_l), ~s_l=0,1$.  Then we 
define the plaquette variable $\sigma_{P} = \mbox{d}s~~ \mbox{mod}~ 2$, 
$(\sigma_P = 0,1)$. The nexus current (or center monopole 
current~\cite{cpvz}) is then defined as 
$^{\star}j=\frac{1}{2}\delta^{\star}\sigma_P$. These currents live on the 
surface of the P-vortex (on the dual 4D lattice) and P-vortex flux goes 
through positive and negative nexuses in alternate order. The important 
characteristic of the cluster of currents is the condensate, $C$, defined 
\cite{polikarpov} as the percolation probability. As it is shown in ref. 
\cite{cpvz} the condensate $C$ of the nexus currents is the order parameter 
for the confinement -- deconfinement phase transition. We found that $C$ is 
nonzero for the gauge copies obtained via RO procedure (when the projected 
Wilson loops have the area law). On the other hand $C$ is zero (in the 
thermodynamic limit $L \rightarrow \infty$) for gauge copies obtained using
LRO procedure (when the projected Wilson loops have no area law). It is
interesting that for RO procedure $C$ seems to scale \cite{bkpv} as the
physical quantity with the dimension $(mass)^4$ . This is illustrated in
Fig.1, where we plot the $\beta$--dependence of the ratio $C/(\sigma_{SU(2)}
a^2)^2$. Thus these new objects might be important degrees of freedom for
the description of the nonperturbative effects.

\begin{figure}
\begin{centering}
\epsfig{figure=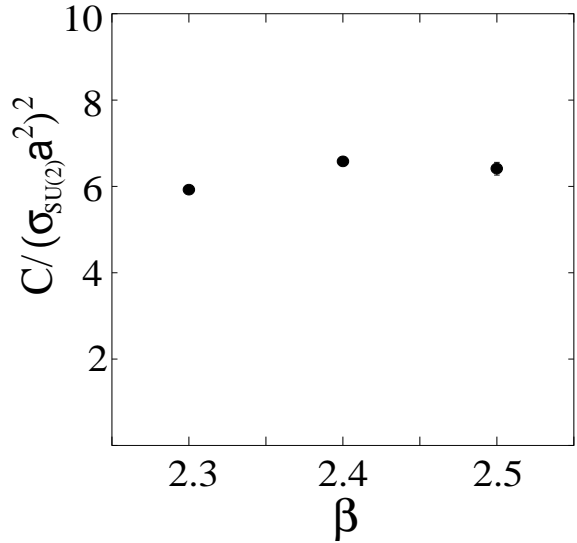,height=7cm,width=7.5cm}
\caption{The $\beta$--dependence of the ratio of the nexus
condensate, $C$, to the $SU(2)$ string tension in lattice units.}
\end{centering}
\end{figure}

\section{Indirect maximal center gauge}
The gauge fixing condition for IMC gauge consists of two steps : \\
Maximal abelian gauge  ($SU(2) \rightarrow U(1)$) is fixed by
solving maximization problem :
$$
\max_{\{g\}} {\{F_1(U^g)\}} ,~~g \in SU(2)/U(1);~~~ F_1(U) = \frac{1}{8 V}
\sum_{n,\mu}  \tr \left( U_{n,\mu}\sigma_3  U^{\dagger}_{n,\mu} \sigma_3
\right)
$$
and $U(1)$ field is extracted: $  U_{n,\mu} =  V_{n,\mu} u_{n,\mu} \, ,~~
u_{n,\mu} \in U(1)$.

\noi
Then $Z(2)$ gauge is fixed by maximizing
$$
\max_{\{\omega\}} {\{F_2(u^{\omega})\}};~~ \omega \in U(1)/Z(2),~~~
F_2(u) = \frac{1}{4 V} \sum_{n,\mu}
\left( \mbox{Re} u_{n,\mu} \right)^2.
$$
The $Z(2)$ field is defined as: $ u_{n,\mu} = Z_{n,\mu}  v_{n,\mu}$.

In this section we discuss our results obtained with two gauge fixing procedures:
\begin{itemize}
\item [--] RO procedure: RO algorithm at  both steps;
\item [--] LRO procedure: LRO algorithm at  the 1st step and RO algorithm  at
the 2nd one;
\end{itemize}
Our results for the ratios $\sigma_{Z(2)}/\sigma_{SU(2)}$
and $2\rho/\sigma_{SU(2)} a^2$  obtained with RO procedure are presented in
the Table 3. These results are in good qualitative and even quantitative
agreement with the results obtained in DMC gauge. The
LRO procedure (as in DMC gauge case) corresponds to 
$\sigma_{Z(2)}=0$. Another important observation is that at the first stage 
(MA gauge) LRO procedure local maxima average $<F_{1,max}>$ is higher than 
 those of RO procedure (see Table 4). Thus the problem discussed in the 
previous section is also relevant for IMC gauge.

\begin{table}[tp]
\caption{The comparison of $\sigma_{Z(2)}$, $\sigma_{SU(2)}$ and $\rho$ for
IMC gauge,  RO gauge fixing procedure. } \label{t1} \vspace{0.5cm}
\setlength{\tabcolsep}{0.55pc}
\begin{centering}
\begin{tabular}{c|ccc|ccc}  \hline
& &$\sigma_{Z(2)}/\sigma_{SU(2)}$&&&$2\rho/(\sigma_{SU(2)} a^2)$& \\ \hline
$N_{cop}$ &$\beta=2.3$&$\beta=2.4$&$\beta=2.5$&$\beta=2.3$&$\beta=2.4$&$\beta=2.5$   \\ \hline
(1,1) &1.03(1)&1.06(1) &1.09(1) &1.51(1) &1.84(1)    &2.17(1)\\
(20,10)&0.89(3) &0.81(2) & 0.86(2)&1.44(1)  &1.69(1)& 1.97(2)\\
($\infty$,$\infty$)&0.81(4) &0.69(3)&0.72(3)&1.40(1)&1.63(1)&1.83(2) \\ \hline
\end{tabular} \end{centering} \end{table}

\begin{table}[tp]
\caption{$<F_{1,max}>$ for IMC gauge obtained with various algorithms}
\label{t1} \vspace{0.5cm}
\setlength{\tabcolsep}{0.55pc}
\begin{centering}
\begin{tabular}{c|ccc}  \hline
&$\beta=2.3$ &$\beta= 2.4$ &$ \beta= 2.5$  \\ \hline
RO($N_{cop}=\infty$)&0.7132(1)&0.7324(3)&0.7509(2) \\
LRO($N_{cop}=\infty$)&0.7134(1)&0.7331(2)& \\
SA($N_{cop}=\infty$) &&0.7337(2)& \\ \hline
\end{tabular} \end{centering} \end{table}

\section{Results of the SA algorithm .}

The simulated annealing algorithm has been applied to the maximal abelian gauge
fixing in \cite{bbms,kerler} and its advantages in reducing the bias due to
Gribov copies effects has been demonstrated. 
In this section we describe our preliminary
results in fixing for both DMC and IMC gauges obtained with SA algorithm.  
For IMC gauge SA algorithm has been applied at  the first step only, while 
RO algorithm has been applied at  the 2nd step. The details of the SA 
algorithm for DMC gauge will be explained elsewhere.

Our first observation is that with SA algorithm it is  possible to reach 
higher local maxima $<F_{max}>$ (for DMC gauge) and $<F_{1,max}>$ (at the 
first step of IMC gauge - MA gauge) than those reached with the LRO 
procedure. This can be seen from the Tables 2 and 4. Our second observation 
is that results obtained with these local maxima are qualitatively similar 
to those obtained with RO procedure, i.e. $\sigma_{Z(2)}$ computed on these 
configurations is nonzero.  The values of $\sigma_{Z(2)}$ and $\rho$ are 
lower than those obtained with RO algorithm.

For SA algorithm applied to DMC gauge
we generated only 3 local maxima per configuration,
since this algorithm is rather time consuming.
Thus we are not able to make extrapolation $N_{cop} \ra \infty$.
Nevertheless the example $N_{cop}=3$ is  instructive. Our results are
$\frac{\sigma_{Z(2)}}{\sigma_{SU(2)}} = 0.69(2), 0.72(2), 0.49(2)$ at
$\beta=2.3,2.4,2.5$ correspondingly.  Comparing with results in Table 1 one
can see that already for $N_{cop}=3$ SA values for $\sigma_{Z2}$ are
lower than values obtained with RO procedure in
$N_{cop}=\infty$ limit. For IMC gauge  with SA algorithm at the first step
we get $\frac{\sigma_{Z(2)}}{\sigma_{SU(2)}} = 0.73(4)$ at $\beta=2.4$
(computations have been made at  this $\beta$ only). This is in agreement
with the result in Table 3 obtained with RO procedure.
It is still possible that employing  SA algorithm  also at  the second  step
of IMC gauge fixing will bring  further decreasing of the projected
string tension.

In \cite{greens4}
it has been argued that low values for $\sigma_{Z2}$
reported in \cite{bkpv} can be due to
finite volume effects
which spoil vortex finding property even on the lattices where
finite volume effects are not visible in the gauge invariant observables.
We made some computations for DMC gauge with RO procedure for lattices
$10^4$ to $20^4$
and found out results in qualitative agreement with fig.~3 of ref.
\cite{greens4}. But the problem is still not settled  since
increasing
of  $\sigma_{Z2}$ with lattice volume can be due to lower value of
local maxima found on larger lattices 
where finding higher maxima becomes too costly,  while
there is clear anticorrelation between $\sigma_{Z2}$  and the value of
$<F_{max}>$ \footnote{We thank Ph. De Forcrand for this remark}.
Moreover  our results with SA algorithm indicate that
these finite volume effects must be much more pronounced if one takes the 
highest maxima available. We made computation in DMC gauge on $L=16$ lattice 
at $\beta=2.4$. We apply the SA algorithm with more updating sweeps than it 
was done for $L=12$ computations at the same $\beta$. We obtained higher 
value $<F_{max}>=0.7784(1)$ and lower value 
$\frac{\sigma_{Z(2)}}{\sigma_{SU(2)}} = 0.64(2)$ than corresponding values 
given in Table 1. The question of the finite volume effects for IMC  gauge 
has not been studied so far.

\section*{Conclusions}

We conclude that DMC and IMC gauges suffer from strong Gribov copies effects
and careful gauge fixing  is necessary to make the bias caused by these
effects reasonably small. This procedure is rather costly.
Another problem of DMC and IMC gauges is that $\sigma_{Z(2)}$ is too
low. It is not clear whether relation $\sigma_{Z(2)} = \sigma_{SU(2)}$
can be valid at large enough lattices. Even if this is the case, the size
of these lattices is enormous.  The alternative gauge, the Laplacian center
gauge \cite{def}  is then more favorable.

\section{Acknowledgments}
V.B. and M.P. are very much thankful to the Organizing Committee
for financial support and for making the meeting very inspiring and
enjoyable.  This study was partially supported by grants RFBR 96-15-96740,
RFBR 99-01230a, INTAS 96-370 and Monbushu grant.

\end{document}